\documentclass[runningheads]{llncs}
\usepackage[T1]{fontenc}

\usepackage{graphicx}

\usepackage{graphicx, hyperref, setspace, amsmath, amssymb, fancyhdr, multicol, parskip, indentfirst, etoolbox, caption, cite, hyperref, xcolor}
\usepackage{booktabs}
\usepackage{multirow}
\usepackage{threeparttable}
\usepackage{tablefootnote}
\usepackage{tabularx}
\usepackage{float}
\usepackage{enumitem}
\begin{document}
\title{Dynamic Context Selection for Retrieval-Augmented Generation: Mitigating Distractors and Positional Bias}

\author{Maya Iratni\orcidID{0000-0002-3567-7885} \and
 Mohand Boughanem\orcidID{0000-0001-7004-0807} \and
Taoufiq Dkaki\orcidID{0000-0003-3962-7663}}
\authorrunning{M. Iratni et al.}

\institute{Institut de Recherche en Informatique de Toulouse (IRIT), Toulouse, France 
\email{\{Malika.Iratni, Mohand.Boughanem, Taoufiq.Dkaki\}@irit.fr}\\}
\maketitle            
\begin{abstract}

Retrieval-Augmented Generation (RAG) enhances language model performance by incorporating external knowledge retrieved from large corpora, which makes it highly suitable for tasks such as open-domain question answering. Standard RAG systems typically rely on a fixed top-k retrieval strategy, which can either miss relevant information or introduce semantically irrelevant passages, known as distractors, that degrade output quality. Additionally, the positioning of retrieved passages within the input context can influence the model’s attention and generation outcomes. Context placed in the middle tends to be overlooked, which is an issue known as the "lost in the middle" phenomenon. In this work, we systematically analyze the impact of distractors on generation quality, and quantify their effects under varying conditions. We also investigate how the position of relevant passages within the context window affects their influence on generation. Building on these insights, we propose a context-size classifier that dynamically predicts the optimal number of documents to retrieve based on query-specific informational needs. We integrate this approach into a full RAG pipeline, and demonstrate improved performance over fixed-k baselines.

\end{abstract}
\section{Introduction}

Retrieval-Augmented Generation (RAG) is an approach that enhances large language models with external knowledge by retrieving relevant passages during inference. In a standard RAG system, the user's query is first encoded, and used to retrieve a ranked list of documents from a large collection \cite{gao2023retrieval} \cite{lewis2020retrieval}. The top-K passages, where {\it k} is a fixed, predetermined value, are then provided to the generative model as additional context, allowing the model to generate a response based on both the query and the retrieved information \cite{lewis2020retrieval} \cite{guu2020retrieval}. This top-K approach has become widely adopted because it is straightforward to implement, and generally provides adequate context for many tasks such as open-domain question answering \cite{gao2023retrieval}. RAG systems often suffer from two critical challenges: (i) misleading irrelevant retrieved documents (distractors), that can distort the generation results \cite{amiraz2025distracting}, and (ii) the ’lost-in-the-middle’ phenomenon. \cite{liu2023lost}. 
A distractor is a retrieved text that appears similar to the query, but is actually semantically irrelevant. Such documents can confuse the generator and reduce output quality by introducing noise into the context, which can mislead the generator and weaken attention over relevant passages \cite{amiraz2025distracting} \cite{jin2024long}. Due to this, using a fixed number of retrieved documents for all queries has significant limitations. A fixed, top-k strategy risks excluding important information when {\it k} is too small or including excessive irrelevant content when {\it k} is too large, making it suboptimal for complex tasks, such as multi-hop QA \cite{su2024dragin}. This balancing problem is directly related to the precision–recall trade-off in retrieval. Increasing {\it k} generally improves recall, so fewer relevant passages are missed. However, it simultaneously deceases the precision by including more irrelevant context, diluting relevant documents and degrading generation quality \cite{jin2024long}. 
An additional challenge in retrieval-augmented generation lies in the positional effect of the retrieved passages within the input sequence. Prior studies have shown that Large Language Models (LLM) tend to prioritize information that appears at the beginning or end of the input sequence, attributing less attention to information placed in the middle. This leads to the "lost in the middle" effect \cite{liu2023lost}, wherein the impact of a relevant document on generation may be diminished, or amplified, depending on its location in the context provided to the generator. This positional sensitivity poses a challenge for retrieval strategies that return many documents, as relevant passages can become buried among distractors and receive less attention. This highlights the importance of the structure of the input, in this case the order in which relevant documents are provided to the generator.
The above points emphasize the fundamental questions of:
\begin{itemize}
    \item \textbf{RQ1:} How much context does a given query need to be effectively answered?
    \item \textbf{RQ2:} Can the impact of distractors be reduced by leveraging the position of relevant context within retrieved documents?
\end{itemize}
 Queries differ in complexity, scope, and information requirements. Some may be answerable with a single passage, while others require broader contextual evidence. This is particularly true in the case of multi-hop question answering. To address this challenges, this work makes the following contributions:
\begin{itemize}
    \item We conduct an empirical study of how distractor passages affect generation quality. Our evaluation quantifies the performance degradation as distractor ratios increase, and identifies the conditions under which distractors most significantly impair model output.
    \item We examine the “lost in the middle” phenomenon by placing relevant passages at the beginning, middle, and end of the model’s input window, to evaluate how positional placement influences the effectiveness of retrieved context for generation within a RAG pipeline.
    \item We introduce a classifier that predicts the optimal number of contexts to retrieve ({\it k}), to dynamically adapt the context length to the specific informational requirements of each query.
    \item We integrate the classifier into a full RAG system and compare its performance against a fixed-k baseline, demonstrating consistent improvements in generation.
\end{itemize}

\section{Related works}
Recent work has explored adaptive retrieval strategies to overcome the limitations of fixed-k retrieval in RAG systems. Jeong et al. (2024) introduce Adaptive-RAG \cite{jeong2024adaptive}, a framework that dynamically selects among non-retrieval, single-step, and multi-step retrieval strategies based on query complexity. Their approach employs a classifier trained to predict query complexity, which guides the selection of an appropriate retrieval depth. 
Sun et al. (2025) introduces DynamicRAG \cite{sun2025dynamicrag}, which adaptively determines both the ranking and the number ({\it k}) of retrieved documents for each query. The core component is a dynamic reranker trained using reinforcement learning, where the quality of LLM-generated responses serves as the reward signal. 
Taguchi et al. propose Adaptive-k \cite{taguchi2025efficient}, a retrieval method that dynamically selects the number of passages to retrieve based on similarity score distribution. Ìn addition to retrieval size, the positional placement of retrieved content has also been shown to influence generation performance. Liu et al. \cite{liu2023lost} demonstrated that large language models often exhibit a position bias, prioritizing information at the beginning and end of the context, and neglecting passages located in the middle. 
In contrast to prior methods, our approach employs a classifier that directly predicts the precise number of documents required for each query. While methods such as Adaptive-RAG rely on classifiers to select among predefined retrieval strategies, our approach explicitly estimates the exact number of documents required for each query.

\section{Preliminary Analysis}
\subsection{Dataset Overview}
We conduct our experiments using the MuSiQue-Ans dataset, a benchmark designed to evaluate complex multi-hop reasoning in open-domain QA systems. The dataset is comprised of queries created through the composition of two to four single-hop queries, combined into a single coherent multi-hop question that requires multiple reasoning steps. The dataset contains a collection of $\approx 22\text{k}$ queries, ($\approx 20\text{k}$ in the training set, and $\approx 2.5\text{k}$ in the dev set), wherein a query can be categorized as either 2-hop, 3-hop, or 4-hop. These ‘hops’ represent the number of reasoning steps required to answer the query. 
Each multi-hop question is paired with a fixed number of ‘gold’ supporting passages, wherin 2-hop questions are provided with two gold documents, 3-hop with three, and 4-hop with four. Table ~\ref{table:table1} depicts the number of queries belonging to each hop type in both the train and dev set.

\noindent
\begin{minipage}{\columnwidth}
\captionof{table}{Number of queries by type} 
\label{table:table1}
\centering
\begin{tabular}{|c|c|c|c|}
    \hline
    \textbf{} & \textbf{2-hop} & \textbf{3-hop} & \textbf{4-hop} \\
    \hline
    Train & 14376 & 4387 & 1175 \\
    Dev & 1252 & 760 & 405\\
    \hline
\end{tabular}
\end{minipage}

\subsection{Exploratory Analysis}
We conducted a series of experiments with the goal of evaluating the impact of distractors, and context positioning on the overall generation performance of a basic RAG system. We examined how three factors influence the answer quality : (i) the omission of gold supporting passages, (ii) the inclusion of distractor documents, and (iii) the position of relevant context in the input sequence.
\begin{figure}[htbp]
  \centering
  \includegraphics[width=\linewidth]{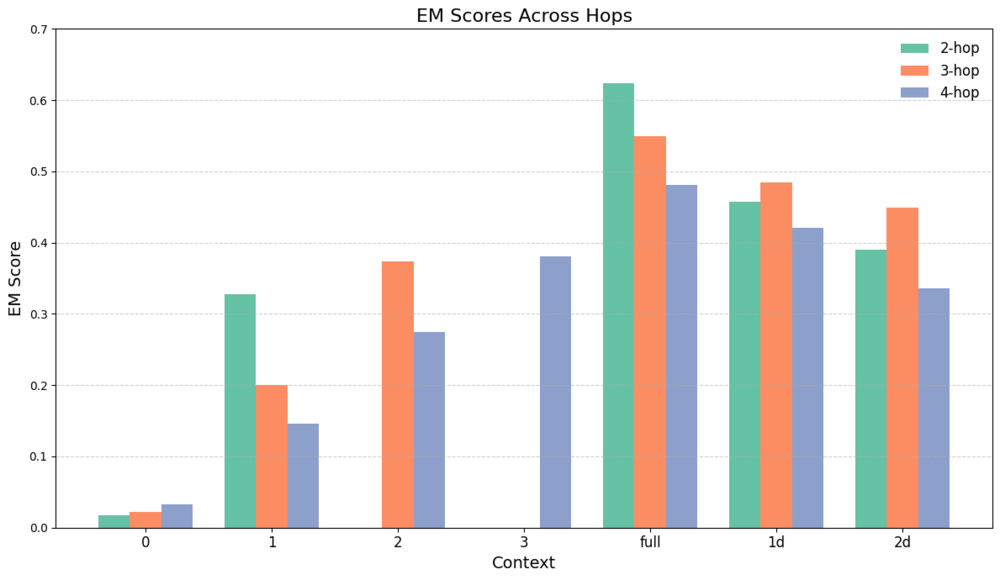}
  \caption{Evaluation of the impact of additional context for 2- to 4-hop and including the presence of distractors (d).}
  \label{fig:three-in-fullwidth}
\end{figure}
\vspace{-10pt}  
\subsubsection{The Impact of Context and Distractors}
The objective of this section is to assess the generation quality of a RAG system under varying levels of supporting context, using the MuSiQue train set. We first isolated 2-hop, 3-hop, and 4-hop questions for separate evaluation. For each case generation performance was iteratively measured by providing the generator with the query alongside: no relevant context, then one, two, three, and four relevant passages iteratively. The goal is to demonstrate how generation quality scales with the amount of relevant contexts provided. 

In addition, we conducted two further iterations on which each query was provided with all its relevant context, followed by adding one, then two distractor passages. The main goal was to simulate retrieval noise, and evaluate the extent to which distractors degrade generation quality.
The results confirm a clear positive correlation between the number of relevant documents and generation quality, but also highlight the destructive impact of distractors.
Figure  \ref{fig:three-in-fullwidth} depicts this finding. 
The results show that distractors cause a significant drop in performance, with a compounding effect as additional distractors are introduced. This is specifically true in the case of 2-hop questions, where a single distractor dropped the performance by more than 26\%. While both 3-hop, and 4-hop queries proved to be more resilient to distractor passages (with a 13.5\% drop for 3-hop, and a 14.4\% drop for 4-hop with a single distractor.), they still experienced a deterioration in generation quality, with a sharper drop when two distractors are added.
\vspace{-0.5em}
\subsubsection{Context Positioning}
This section examines the effect of the position of relevant passages on generation quality. For each query (2-hop, 3-hop, and 4-hop), we retrieved 5 passages, including distractors and relevant passages. The positions varied according to three configurations: relevant passages placed at the beginning, placed in the middle, placed at the end of the sequence.
As shown in Table \ref{table:rank_order}, placing relevant passages at the end led to the best performance, while placing them in the middle resulted in the lowest generation quality, supporting the 'lost in the middle' phenomenon.
These results are consistent with known input position biases in LLMs, which tend to prioritize information appearing at the extremities of the input.
The end-position advantage suggests models prioritize more recent content.
These findings highlight the importance of context placement in RAG systems. When input space is limited or distractors exist, placing relevant content near the end improves generation.

\noindent
\begin{minipage}{\columnwidth}
\captionof{table}{Relevant context position} 
\label{table:rank_order}
\centering
\begin{tabular}{lccc}
    \hline
    \textbf{} & \textbf{Beginning} & \textbf{Middle} & \textbf{End} \\
    \hline
    Exact Match & 0.5447 & 0.5391 & 0.5567 \\
    F1 score & 0.6212 & 0.6126 & 0.6361 \\
    \hline
\end{tabular}
\end{minipage}

\section{Proposed Approach}
\begin{figure*}[t]
\centering
\includegraphics[width=\textwidth]{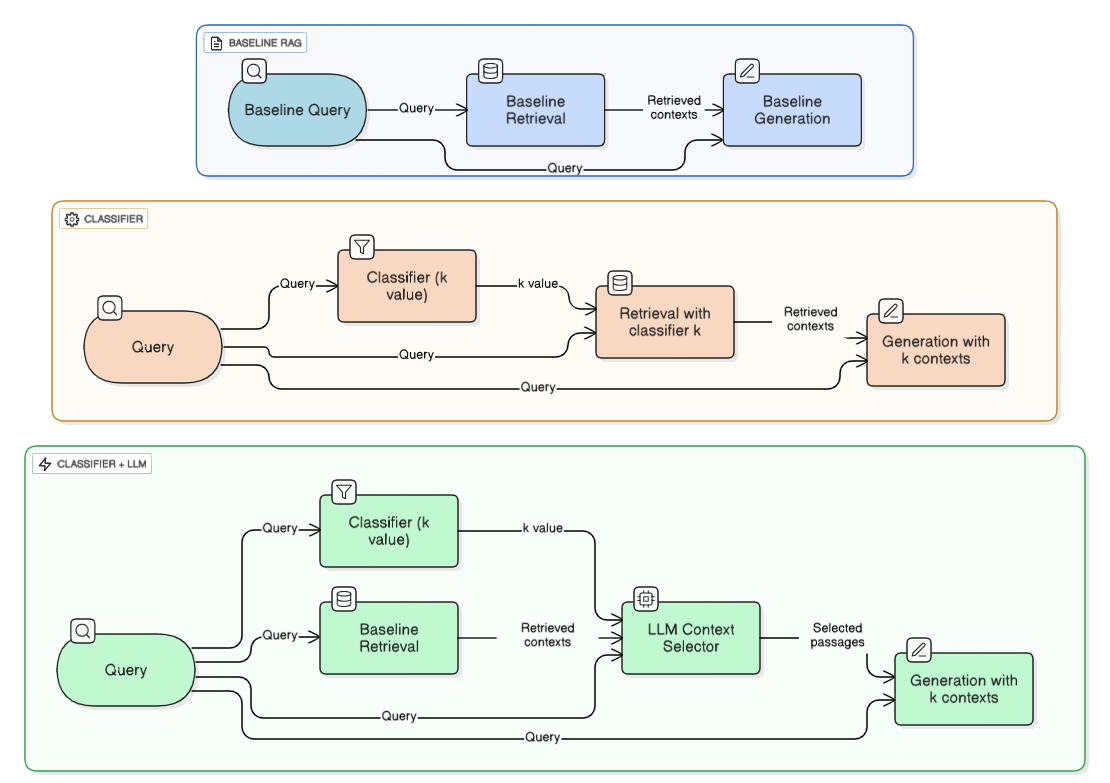}
\caption{Diagram of Baseline Pipeline, Classifier-k Pipeline, and Classifier-k+LLM pipeline}
\label{fig:diagram}
\end{figure*}

Based on our preliminary analysis, we propose an enhanced RAG system designed to improve generation quality for multi-hop QA by reducing the impact of distractors. This section presents the architecture and components of our system which extends the standard RAG pipeline with two extensions designed to reduce noise from irrelevant retrieved passages. 
Figure ~\ref{fig:diagram} illustrates the overall system architecture, contrasting the baseline RAG pipeline with our proposed extensions: including the dynamic-k classifier and the LLM-based reranker.

\subsection{Baseline Pipeline}
Our starting point is a standard RAG setup consisting of two main modules:
\begin{itemize}[noitemsep, topsep=0pt]
\itemsep0em 
    \item \textbf{Retriever:} Given an input query $q$, the retriever searches through a corpus and returns the top-$k$ passages most relevant to $q$.
    \item \textbf{Generator:} The retrieved passages are concatenated with $q$ and fed into a generative model, which produces the final answer.
\end{itemize}

Given a query $q$, the retriever fetches $k$ passages $P_k = \{p_1, p_2, \ldots, p_k\}$, and the generator produces output:
$y = \text{Generator}(q, P_k)$

This architecture provides strong end-to-end performance, but uses a fixed $k$ for all queries, which can lead to over or under retrieval depending on query complexity.

\subsection{Classifier-k Pipeline}
The first major modification to this pipeline is the integration of a query-specific $k$ predictor. We introduce a classifier to estimate the optimal number of context ($k$) that should be retrieved for each query. The classifier takes as input a given query and outputs an integer value corresponding to the number of contexts to retrieve. This value is used to dynamically adjust the number of documents retrieved for each query.

The pipeline is thus modified as follows:
\begin{enumerate}[noitemsep, topsep=0pt]
    \item The classifier processes query $q$ to predict retrieval count: 
    \[
    k_{\text{pred}} = \text{Classifier}(q)
    \]
    \item The retriever fetches $k_{\text{pred}}$ passages.
    \item The generator produces output: 
    \[
    y = \text{Generator}(q, P_{k_\text{pred}})
    \]
\end{enumerate}

\subsection{Classifier-LLM Pipeline}
The second major modification to the pipeline augments the system an LLM module for reranking. In this pipeline, we first retrieve a fixed-k number of candidate contexts, as was done in the basic retrieval pipeline. The classifier-predicted $k$ value, the original query, and the top $k_{\text{fixed}}$ retrieved contexts are passed as input to an LLM, which will act as an additional reranker. The LLM is prompted to select the top classifier-k passages from the candidate set based on their relevance to the query. These selected passages are then used as the context for the final generation step.

The pipeline is thus modified as follows:
\begin{enumerate}[noitemsep, topsep=0pt]
    \item The classifier processes query $q$ to predict retrieval count:
    \[
    k_{\text{pred}} = \text{Classifier}(q)
    \]
    \item The retriever fetches $k_{\text{fixed}}$ passages (basic retrieval).
    \item The LLM reranker filters passages:
    \[
    P_{\text{filtered}} = \text{LLM\_Reranker}(q, k_{\text{pred}}, P_{\text{fixed}})
    \]
    \[
    |P_{\text{filtered}}| = k_{\text{pred}}
    \]
    \item The generator produces output:
 $ y = \text{Generator}(q, P_{\text{filtered}})$
 
\end{enumerate}
\section{Experimental setup}
To evaluate our proposed approach, we conducted experiments using a variety of retrievers on three datasets, using the three pipelines. We describe the relevant configurations for each setup below. 

\subsection{Retrievers}
\begin{itemize}
    \item \textbf{BM25:} A sparse retrieval with scores computed by the BM25 model.
    \item \textbf{Dense Retrieval}: A vector-based retrieval system using cosine similarity. Passages were embedded using the \texttt{all-MiniLM-L6-v2} model from SentenceTransformers, and the Chromadb vector store.
    \item \textbf{ColBERT + MonoT5}: A two-stage retrieval pipeline using ColBERT and a monoT5 reranker.
    \item \textbf{ColBERT + BAAI/bge-reranker-large}: An alternative configuration using ColBERT paired with the BGE-large reranker.
\end{itemize}

Both two-stage configurations included the retrieval of top-50 candidates with ColBERT and top-$k$ reranking with either MonoT5 or BGE-Reranker.

\subsection{Dataset and Evaluation Metrics}
\begin{itemize}
    \item \textbf{MuSiQue: \cite{trivedi2022musique}} We aggregated all unique passages from the dev set of the MuSiQue-Ans dataset to form our retrieval corpus, yeilding $\approx 22\text{k}$ passages.
    \item \textbf{2WikiMultihopQA: \cite{ho2020constructing}} No retrieval component was applied for this dataset. Instead, a reranker was employed to reorder the 10 candidate texts per query.
 
    \item \textbf {MultihopRAG \cite{tang2024multihop}:} This dataset includes a corpus of 603 full-length texts. Basic chunking was applied. An 80/20 split was used, with 80\% of the queries allocated for training the classifier and 20\% reserved for evaluation. 
    \item \textbf{Evaluation:} Retrieval quality was measured using precision and recall. Generation quality was measured using Exact Match (EM) and F1-score (F1).
\end{itemize}

\subsection{Classifier}
To dynamically predict the number of passages ($k$) to retrieve per query, we fine-tuned a RoBERTa model on the training collection of all three datasets. The classifier was trained as a  multi-class task, where the input is a question and the target is its annotated hop type: 2-hop, 3-hop, or 4-hop. The output class is then mapped to a corresponding $k$ value for retrieval. The classifier was evaluated on all three datasets, and was evaluated to be 87.3\% accurate using exact match. An alternate classifier was trained using only the MuSiQue dataset, to evaluate if the performance is affected significantly when trained on three different datasets. The classfier trained on MuSiQue only scored a 77.8\% accuracy on MuSiQue queries, while the classifier trained on all three datasets scored 76.5\%, which does not indicate a substantial loss in performance.

\begin{itemize}
\itemsep0em
    \item \textbf{Model architecture:} We used the \texttt{roberta-base} model from HuggingFace, with a classification head on top.
    \item \textbf{Input:} Natural language question.
    \item \textbf{Output:} A categorical label indicating the number of reasoning hops required.
\end{itemize}

\textbf{Training Configuration:} The model was fine-tuned for 5 epochs using the AdamW optimizer with a learning rate of $2 \times 10^{-5}$ and linear learning rate scheduling. In order to address class imbalance in the hop type distribution, which is severely skewed in the favor of 2-hop questions, we implemented a \texttt{WeightedRandomSampler} that assigns inverse frequency weights to each class during training. The dataset was split into 70\% training, 15\% validation, and 15\% test sets using stratified sampling to maintain class proportions across splits. Training was conducted with a batch size of 16, and the best model was selected based on validation accuracy.

\subsection{LLM Reranker}
For the document selection component, we employed Mistral Nemo Instruct (12.2B parameters) with zero-shot prompting. The prompting strategy instructed the model to: (1) Analyze the given query and retrieved passages, (2) select exactly $k$ passages (predicted by the classifier) based on relevance to the query, and (3) return relevant passages. 
 The prompt is as follows: "\emph{Given the following query and passages, rank the passages (by their ID numbers) that are most relevant to answering the query. Return the {predicted-k} most relevant passage IDs in a Python list.}"

\subsection{Generator}

The final generation stage utilized Flan-T5-XL as the text generation model, which processes the selected passages alongside the original query to produce the final response.

\section{Results and Discussion}

\subsection{Baseline vs Classifier-k Pipeline}
This section compares the performances of the Baseline and the Classifier pipelines. The fixed-k value for the Baseline Pipeline was selected to be a standard value of five (k=5). 
Table ~\ref{table:metrics-results} shows the generation results for each configuration. The evaluation shows that the Baseline outperformed the Classifier pipeline (Baseline vs. Calssifier rows) in both the MuSiQue and 2WikiMultiHopQA datasets, with all retrieval configurations.
Before considering alternative explanations, we must ask whether the issue stems from the classifier, or not. {\bf If it always returned the correct $k$ value, would the results outperform the baseline?} As is depicted in Table ~\ref{table:metrics-results}, the fixed k (Baseline) approach outperforms the classifier-k approach, even when the classifier is ideal. The result is shown in the Ideal Classifier row, using the dense retrieval method, compared the the Baseline.
The retrieval results (not listed in the paper) show that while precision scores were increased with the classifier-k retrieval, the recall significantly decreases. This means that while the approach may be reducing the number of distractor documents, it cuts-off relevant contexts, which decreases generation performance. If a relevant context is found in lower positions, but only the top two contexts are selected, that relevant context is lost. 
Thus, the issue likely stems from retrieval quality rather than the classifier, highlighting the need for a stronger reranker.


%
\begin{table}[t!]
\centering
\begin{threeparttable}
\setlength{\abovecaptionskip}{2pt}
\setlength{\belowcaptionskip}{2pt}
\caption{Evaluation Results}
\label{table:metrics-results}
\centering
\renewcommand{\arraystretch}{1.2}
\begin{tabularx}{\textwidth}{lXcccccc}
\toprule
\multirow{2}{*}{Retrieval} & \multirow{2}{*}{Pipeline} 
& \multicolumn{2}{c}{MuSiQue} & \multicolumn{2}{c}{MultihopRAG} & \multicolumn{2}{c}{2WikiQA} \\
\cmidrule(lr){3-4} \cmidrule(lr){5-6} \cmidrule(lr){7-8}
& & EM & F1 & EM & F1 & EM & F1 \\
\midrule
\multirow{3}{*}{BM25} 
& Baseline        & 0.060 & 0.105 & 0.499 & 0.550 & - & - \\
& Classifier      & 0.046 & 0.086 & 0.501 & 0.549 & - & - \\
& Classifier+LLM  & \textbf{0.063} & \textbf{0.107} & \textbf{0.530} & \textbf{0.585} & - & - \\
\midrule
\multirow{3}{*}{Dense} 
& Baseline        & 0.141 & 0.221 & 0.567 & 0.612 & - & - \\
& Classifier      & 0.124 & 0.195 & 0.594 & 0.642 & - & - \\
& Classifier+LLM  & \textbf{0.153} & \textbf{0.232} & \textbf{0.597} & \textbf{0.648} & - & - \\
\midrule
\multirow{3}{*}{MonoT5} 
& Baseline        & 0.171 & 0.252 & 0.594 & 0.626 & - & - \\
& Classifier      & 0.159 & 0.239 & 0.621 & 0.653 & - & - \\
& Classifier+LLM  & \textbf{0.195} & \textbf{0.277} & \textbf{0.625} & \textbf{0.659} & - & - \\
\midrule
\multirow{5}{*}{BGE} 
& Baseline        & 0.177 & 0.260 & 0.601 & 0.645 & 0.488 & 0.576 \\
& Classifier      & 0.162 & 0.243 & 0.619 & 0.655 & 0.486 & 0.572 \\
& Classifier+LLM  & 0.199* & 0.283* & \textbf{0.625}* & 0.666* & 0.529* & 0.625* \\
& Classifier+LLM (Structured)$^1$  & \textbf{0.202}* & \textbf{0.291}* & \textbf{0.625}* & \textbf{0.672}* & \textbf{0.531}* & \textbf{0.629}* \\
& Control Pipeline$^4$  & 0.171 & 0.212 & - & - & - & - \\
\midrule
\multirow{2}{*}{Oracle Study} 
& Ideal Classifier$^2$       & 0.128 & 0.200 & - & - & - & - \\
& Ideal Reranker$^3$        & 0.216 & 0.303 & - & - & - & - \\
\bottomrule
\end{tabularx}
\begin{tablenotes}
\small
\item[$^1$] The input was reordered so that the most relevant texts are at the end of the input.
\item[$^2$] Comparison of the Baseline pipeline(k=5) and the Classifier pipeline with the optimal k under dense retrieval.
\item[$^3$] Using the BGE retrieval method.
\item[$^4$] In the control pipeline, the reranker LLM was given the candidate texts and tasked with selecting the optimal number independently of the classifier.
\item[$^*$] These measurements were compared against the results obtained using the BGE retriever and Baseline pipeline configuration using a paired t-test. The results show a p-value of <0.01 for all three datasets.
\end{tablenotes}
\end{threeparttable}
\end{table}

\subsection{Oracle Study}
\subsubsection{Ideal Retriever}
The previous experiment raises the question, if the retrieval module is ideal, will the classifier improve the generation performance?
This section simulates an ideal retriever. This means that in the case of the baseline, the fixed-k retrieved contexts will contain all relevant contexts. However, this means that it will also include extra distractor passages. For example, a 2-hop question would have five retrieved contexts, two of which are relevant, and three of which are distractors. 
In the classifier pipeline, the classifier determines the $k$. However, the classifier is not guaranteed to be accurate each time. For example, it could incorrectly label a 3-hop query as 4-hop, wherein all three relevant documents are retrieved, as well as one distractor document. Although, if the classifier mislabels a 3-hop question as 2-hop, only two relevant contexts are be retrieved, and one would be missing. This experiment was conducted four times with the baseline pipeline with varying fixed-k values. The experiment was also conducted with the classifier-k, and with the ideal k value. Table ~\ref{table:k-analysis} shows the results of this experiment.
The best performing setup is with the ideal k, followed by the classifier-k. This shows that with an ideal retriever, the classifier outperforms the fixed-k baseline.

\noindent
\begin{minipage}{\columnwidth}
\setlength{\abovecaptionskip}{3pt}
\setlength{\belowcaptionskip}{3pt}
\captionof{table}{Generation performance with fixed-k, classifier-k, and ideal-k using ideal retrieval}
\label{table:k-analysis}
\centering

\setlength{\tabcolsep}{10pt}
\renewcommand{\arraystretch}{1}
\begin{tabular}{lcc}
\toprule
\textbf{Setting} & \textbf{EM} & \textbf{F1} \\
\midrule
$k=2$        & 0.475 & 0.621 \\
$k=3$        & 0.520 & 0.667 \\
$k=4$        & 0.523 & 0.673 \\
$k=5$        & 0.481 & 0.638 \\
Classifier-k & 0.535 & 0.682 \\
Ideal-k      & 0.564 & 0.705 \\

\bottomrule
\end{tabular}
\end{minipage}

\subsubsection{Ideal Reranker}
In this part, the retrieval is performed for both pipelines (Baseline and Classifier) using a fixed k=5. We then introduce an ‘ideal’ reranking module that reorders the contexts such that relevant contexts appear at the top of the list, followed by any distractors that were retrieved.
For example, if the classifier predicts a question to be 3-hop (k=3), the model selects the first three contexts after ideal reranking, ensuring that if relevant contexts exist in the top five, they are prioritized. Table ~\ref{table:metrics-results} shows the results of this simulation, in the Ideal Reranker row.
The results show that an ideal reranking module with the classifier produces better results than the fixed-k pipeline. Although both pipelines end up with the same number of relevant contexts, the increased presence of distractors in the fixed-k approach decreases the generation performance. 

\subsection{Classifier-LLM Pipeline}
This model replicates the ideal reranker experiment using an LLM  to perform additional reranking. Retrieval is first performed with a fixed $k=5$ and generation follows as in standard fixed-$k$ pipeline. In the classifier+LLM pipeline, the LLM receives the classifer-k value, the five retrieved contexts, and the query. The LLM is prompted to rank the contexts by relevance, and select the top classifier-$k$ contexts. The selected contexts, and the query are then provided to the generator. 
We also evaluated a control pipeline without the classifier, in which the LLM receives the five contexts, and tasked to determine the optimal $k$ itself, and to select the most relevant documents for generator. Table ~\ref{table:metrics-results} reports the results of the Control Pipeline and compares them with the Baseline and the Classifier-LLM pipelines accross each retrieval configuration.
The results show that the Classifier-LLM configuration outperformed the baseline with every retrieval method.
In addition, we applied the concept of context positioning introduced in Section III. Passages judged most relevant by the LLM were placed at the end of the context fed to the generator. As shown in the table in the Classifier-LLM (Structured) part, this modification further improved performance.
To evaluate the significance of the performance difference between our model and the baseline, we conducted a paired t-test between the Baseline and Classifier-LLM pipeline in the BGE retrieval configuration, which yielded the best results. The test yielded a p-value <0.01 on all three datasets, indicating that the improvement achieved by our model is statistically significant.
\vspace{-8pt}
\subsubsection{Comparative Analysis}
We compare the performance of our model against the Adaptive-Rag \cite{jeong2024adaptive} results obtained in their paper, which addresses a related task using a different retrieval method. Results against our Classifier-LLM pipeline (BGE retriever) were comparable on the MuSiQue dataset (with 23.6 against our 20.2 using EM), while our model outperformed on the 2WikiMultihopQA dataset (with 40.6 against our 53.1 using EM). We did not compare our model directly against DynamicRAG \cite{sun2025dynamicrag} as their evaluation was conducted on a different set of datasets, and their approach requires computationally intensive training procedures that are not directly comparable to our setting.

\section{Conclusion}
In this work, we examined two key limitations of standard RAG systems: the adverse impact of distractor passages and the positional bias introduced by the “lost in the middle” effect. 
Using the MuSiQue-Ans dataset, we showed that both distractors and suboptimal passage placement substantially reduce generation quality, particularly for multi-hop queries.
To adress these issues, we introduced a dynamic context selection framework that combines a query-specific classifier with an LLM-based reranking  to dynamically determine the optimal number of documents to retrieve.
Empirical results indicate that while classifier-only retrieval can lower recall, integrating classifier predictions with LLM reranking yields significant improvements in exact match and F1.
 Moreover, positioning the most relevant passages at the end of the context sequence yields additional gains, reinforcing the importance of input structure in RAG pipelines. These findings highlight the need for more dynamic context selection methods.
\vspace{-5pt}
\begin{credits}
\subsubsection{\discintname}
The authors have no competing interests to declare that are
relevant to the content of this article. 
\end{credits}

\bibliographystyle{abbrv}
\bibliography{ref.bib}

\end{document}